\documentclass[superscriptaddress,twocolumn,aps,longbibliography]{revtex4-2}
\usepackage{times,xspace}
\usepackage{amsbsy,amssymb,amsmath,bm}
\usepackage{graphicx,color,epsfig}
\usepackage{fancyhdr}
\usepackage{soul}
\usepackage[colorlinks=true,
linkcolor=blue,
urlcolor=blue,
citecolor=blue]{hyperref}
\usepackage{float}

\begin{document}
\title{Electronic Raman scattering of antiferromagnetic excitonic insulators}

\author{Hidemaro Suwa}
\email{suwamaro@phys.s.u-tokyo.ac.jp}
\affiliation{Department of Physics, The University of Tokyo, Tokyo 113-0033, Japan}
\author{Shang-Shun Zhang}
\affiliation{Department of Physics and Astronomy, The University of Tennessee,
Knoxville, TN 37996, USA}
\author{Cristian D. Batista}
\affiliation{Department of Physics and Astronomy, The University of Tennessee,
Knoxville, TN 37996, USA}
\affiliation{Quantum Condensed Matter Division and Shull-Wollan Center, Oak Ridge National Laboratory, Oak Ridge, TN 37831, USA}

\date{\today}

\begin{abstract}
The excitonic insulator, a quantum mechanical state arising from exciton condensation, was proposed theoretically many years ago but has yet to be experimentally confirmed. The discovery of correlated transition metal oxides based on $4d$ and $5d$ elements, where the on-site Coulomb repulsion is comparable to the dominant hopping amplitude, presents a unique opportunity to study exciton condensation. 
By constructing an effective mean field Raman operator for the Hubbard model, we derive the low-energy electronic Raman scattering cross section, demonstrating Raman spectroscopy as a powerful tool for detecting exciton condensation. 
Here, we demonstrate that Raman scattering directly reveals exciton condensation in the bilayer iridate Sr$_3$Ir$_2$O$_7$ under pressure.
\end{abstract}

\maketitle

The electron Coulomb interaction can lead to the formation of excitons by binding electrons and holes together. The condensation of these excitons leads to long-range order of degrees of freedom dictated by the quantum numbers of the condensing mode.   For instance, charge density waves emerge from the condensation of excitons in the singlet $S=0$ channel, while spin density waves arise from condensing $S=1$ excitons. When the condensed state possesses a finite charge gap, it is referred to as an excitonic insulator.~\cite{Jerome1967excitonic,Halperin1968possible}. 

Excitonic insulator states are thought to exist in materials such as TmSe$_{0.45}$Te$_{0.55}$~\cite{Bucher1991excitonic}, 1$T$-TiSe$_2$~\cite{Cercellier2007}, Ta$_2$NiSe$_5$~\cite{Wakisaka2009excitonic}, and LaCoO$_3$~\cite{Ikeda2023}.
Various experimental results, including the opening of a band gap at low temperatures~\cite{Lu2017,Larkin2017}, the flattening of angle-resolved photoemission spectroscopy (ARPES) spectra~\cite{Wakisaka2009excitonic,Seki2014,Fukutani2021}, and critical fluctuations observed in Raman scattering~\cite{Kim2021,Volkov2021}, support the excitonic insulator state, but these findings are indirect. The excitonic insulator transition can be accompanied by a structural phase transition, and the electron-lattice coupling may be the primary driving force of the transition~\cite{Watson2020,Baldini2023}. 
Conclusive experimental evidence distinguishing whether the transition is primarily electronic rather than structural changes has yet to be confirmed.

Recently, bilayer 4$d$- and 5$d$-orbital Mott systems, such as ruthenates and iridates~\cite{Rau2016,Cao2018,Hao2018,Yang2022,Yang2023}, in the intermediate coupling regime have become excellent candidates for realizing the exciton condensation~\cite{Kunes2015,Mazzone2022,Okuma2024}.
In these materials, the low-energy physics of these compounds can be described by a Hubbard model with an intra-atomic Coulomb repulsion $U$ comparable to the dominant hopping amplitude $t$. The charge gap in these systems can be comparable to the magnetic energy scale (single-magnon bandwidth), 
leading to enhanced charge-spin coupling and consequently a rich spectrum of excitations. However, the experimental identification of these excitations sometimes remain elusive due to their strongly correlated nature.

 For example, the bilayer compound  Sr$_3$Ir$_2$O$_7$ can be modeled by a single orbital Hubbard model of the $J_{\rm eff}=1/2$ orbital, emerging from the $5d^5$ electron configuration in the presence of strong spin-orbit coupling~\cite{Carter2013microscopic,Suwa2021,Suwa2023}. This material has a longitudinal magnetic mode, observed in resonant inelastic X-ray scattering (RIXS) as evidence of an antiferromagnetic excitonic insulator state~\cite{Mazzone2022}. 
The Raman spectrum of Sr$_3$Ir$_2$O$_7$ reveals intriguing low-energy electronic excitations~\cite{Gretarsson2016two,Zhang2019,Li2020symmetry}, with the lowest frequency peak initially identified as a one-magnon mode~\cite{Zhang2019}.
However, subsequent symmetry analysis has refuted this identification, recognizing it instead as a two-magnon mode~\cite{Li2020symmetry}. Moreover, the lowest energy mode exhibits significant softening under hydrostatic pressure, suggesting a quantum phase transition (QPT) to a disordered phase~\cite{Li2013, Zhang2019, Zhao2014, Zocco2014, Ding2016, Donnerer2018}. The origin and nature of this QPT, as well as the lowest frequency Raman peak,  remain poorly understood. Theoretical calculations of the electronic Raman response using the Hubbard model face challenges due to the complexity of the scattering cross-section away from the strong coupling limit $U \gg |t|$. 

In this paper, we extend the low-energy theory of magnetic Raman scattering ~\cite{ShastryS1990,ShastryS1991} of a half-filled two-band Hubbard model to the intermediate coupling regime and provide a theoretical explanation for the experimental observations in the intermediate coupling material Sr$_3$Ir$_2$O$_7$. 
A key distinction from the strong-coupling regime is the presence of charge fluctuations (particle-hole continuum) at energies comparable to those of magnetic excitations.
After introducing a mean-field Raman operator, we compute the low-energy electronic Raman spectrum of Sr$_3$Ir$_2$O$_7$ and identify the lowest energy peak as the longitudinal mode.
We also investigate the pressure dependence of the Raman spectrum, demonstrating that the softening of the longitudinal mode, a fingerprint of the excitonic insulator, aligns with experimental observations. 
Because neutron absorption by iridium hinders a precise study of neutron scattering and RIXS has a  limited energy resolution,  Raman scattering becomes the best-suited probe for detecting exciton condensation. Furthermore, we suggest that the BCS-BEC crossover can be controlled by anisotropic pressure.

\textit{Model.}---We model the bilayer iridate Sr$_3$Ir$_2$O$_7$ using an effective $J=1/2$ single-orbital half-filled bilayer Hubbard Hamiltonian~\cite{Carter2013microscopic,Suwa2021} $\mathcal{H}=-\mathcal{H}_{{\rm K}}+\mathcal{H}_{{\rm I}}$, with $\mathcal{H}_{{\rm I}}\!=U\sum_{\bm{r}}n_{\bm{r}{\uparrow}}n_{\bm{r}{\downarrow}}$
and 
\begin{equation}
\mathcal{H}_{{\rm K}} = \sum_{{\bm{r}},\bm{\delta}_{a}}t_{a}{\bm{c}}_{\bm{r}}^{\dagger}{\bm{c}}_{{\bm{r}}+{\bm{\delta}}_{a}}+t_{z}\sum_{{\bm{r}}_{\bot}}{\bm{c}}_{({\bm{r}}_{\bot},1)}^{\dagger}e^{i\frac{\alpha}{2}\epsilon_{\bm{r}}\sigma_{z}}{\bm{c}}_{({\bm{r}}_{\bot},2)}+{\rm H.c.},
\end{equation}
where 
$t_a \in\mathbb{R}$ ($a$ = 1, 2)
are the nearest- and next-nearest-neighbor hopping amplitudes within the square lattice of each Ir layer, 
$t_z \in\mathbb{R}$ is the hopping amplitude on the vertical bonds between the two layers, and $\sigma_z$ is the Pauli matrix. 
The overall phase was chosen to gauge away the phase of 
$t_a$.
The operator $ c_{\bm{r}}^{\dagger}$ = [$ c_{\bm{r}\uparrow}^{\dagger}$, $c_{\bm{r}\downarrow}^{\dagger}$] creates the 
two-component spinor of the electron field at $\bm{r} = (\bm{r}_\bot, l)$ with $l = 1, 2$ denoting the layer index and $\bm{r}_\bot = r_1\bm{a}_1 + r_2\bm{a}_2$. 
Here, the primitive in-plane lattice vectors are denoted by $\bm{a}_1$ and $\bm{a}_2$, and the directed neighboring bonds are represented by $\bm{\delta}_1=\bm{a}_1,\bm{a}_2$ and $\bm{\delta}_2=\bm{a}_1 \pm \bm{a}_2$. 
In the interaction term $H_{\rm I}$, $U$ is the effective on-site Coulomb interaction, and $n_{\bm{r}\sigma}$ is the number operator of spin-$\sigma$ electrons at $\bm{r}$.
The sign $\epsilon_{\bm{r}}$ of the spin-dependent hopping term takes the values $\pm1$ depending on which sublattice ($\mathcal{A}$ or $\mathcal{B}$) of the bipartite bilayer system $\bm{r}$ points to. 
The key phase $\alpha$ arises from hopping matrix elements between $d_{xz}$ and $d_{yz}$ orbitals, which are allowed to be nonzero by the staggered octahedral rotation in the unit cell~\cite{Carter2013microscopic}.
The system has an easy
$z$-axis spin anisotropy for $\alpha\neq0$, and the ground state
can have Néel ordering, $\langle S_{\bm{r}}^{\mu}\rangle=(-1)^{\gamma_{\bm{r}}}M\delta_{\mu z}$,
where $\gamma_{\bm{r}}=(1+\epsilon_{\bm{r}})/2$, ${S}_{\bm{r}}^{\mu}={1/2}c_{\bm{r}}^{\dagger}\sigma^{\mu}c_{\bm{r}}$
$(\mu=x,y,z)$ and $M$ is the magnetization.

\textit{Raman scattering}.--- To compute the electronic (non-phonon) contribution to Raman scattering we must start from the  interaction between electrons and light can be added to the Hubbard model by using the Peierls construction. 
Assuming that the spatial variation of the vector potential $\bm{A}$ is much slower than the lattice spacing scale and expanding in powers of $\bm{A}$ up to the second order, results in the following scattering Hamiltonian~\cite{ShastryS1990,ShastryS1991}:
\begin{equation}
\begin{split}
\mathcal{H}_{{\rm int}} &=-\frac{e}{\hbar c}\sum_{\bm{q}}\bm{j}(\bm{q})\cdot\bm{A}(-\bm{q})\\
&+\frac{e^{2}}{2\hbar^{2}c^{2}}\sum_{\bm{q}_{1}\bm{q}_{2}\mu\nu}A_{\mu}(-\bm{q}_{1})\tau_{\mu\nu}(\bm{q}_{1}+\bm{q}_{2})A_{\nu}(-\bm{q}_{2}),
\end{split}
\end{equation}
where $\bm{j}(\bm{q})$ is the Fourier transform of the current density operator, $\tau$ is the second rank stress tensor,
$e$ is the electron charge, $\hbar$ is the Planck constant, and $c$ is the speed of light. 
We consider the in-plane components: $\mu,\nu=x,y$. The cross section is determined by the transition probability rate $R$ obtained from Fermi's golden rule:
\begin{align}
R^{\bm{e}^i \bm{e}^f} \!\!(\bm{q},\Omega)
= C  e^{-\beta\epsilon_{\eta}} \rvert {\bm e}^i_{\mu} {\bm e}^f_{\nu} \langle \gamma \rvert \hat{M}^{\mu\nu}(\bm{q})\rvert \eta\rangle\rvert^{2}
\delta(\epsilon_{\gamma}-\epsilon_{\eta}-\Omega),    
\nonumber \\
\label{eq:cross-section}
\end{align}
where $\rvert \eta\rangle$ $(\rvert \gamma\rangle)$ is the initial (final) eigenstate of the system with eigenenergy $\epsilon_{\eta(\gamma)}$ and we are adopting the convention of summation of repeated Greek indices. The energy and momentum of the initial (final) state of a photon are denoted by $\omega_{i(f)}$ and $\bm{k}_{i(f)}$, respectively, $\bm{q}=\bm{k}_{i}-\bm{k}_{f}$, $\Omega=\omega_{i}-\omega_{f}$, and the proportionality constant is
$C=2\pi g^2_i g^2_f$, with $g(\bm{q})=(hc^{2}/V\omega_{\bm{q}})^{1/2}$, $g_i=g(\bm{k}_{i})$, and $g_f=g(\bm{k}_{f})$. The scattering operator reads $\hat{M}^{\mu\nu}(\bm{q})=\hat{M}^{\mu\nu}_{N}(\bm{q})+\hat{M}^{\mu\nu}_{R}(\bm{q})$, with the non-resonant part $\hat{M}_{N}^{\mu\nu}(\bm{q}) = \tau_{\mu\nu}(\bm{q})$ and the resonant contributions
\begin{equation}
 \hat{M}_{R}^{\mu\nu}(\bm{q}) 
 = j_{\nu}(\bm{k}_{f})\hat{G}_{\epsilon_{\eta}+\omega_{i}} j_{\mu}(-\bm{k}_{i}) + j_{\mu}(-\bm{k}_{i}) \hat{G}_{\epsilon_{\eta}-\omega_{f}} j_{\nu}(\bm{k}_{f}),
\label{eq:M_R}
\end{equation}
with $\hat{G}_E = ( \hat{\mathcal H} -E)^{-1} $, 
and $j_{\mu}(\bm{k})$ is the $\mu$-component of $\bm{j}(\bm{k})$.
The nonresonant part is expressed as a bilinear form of the fermionic operators, making it relatively straightforward to calculate. In contrast, calculating the resonant part is generally complex due to the presence of the resolvent. In the strong coupling regime, a perturbation expansion in terms of $t/(U-\omega_i)$ yields the  Fleury-Loudon scattering Hamiltonian~\cite{ShastryS1990,ShastryS1991,Fleury68,Elliott63}, which is commonly used for spin systems~\cite{Devereaux2007}. However, this perturbative approach fails in intermediate coupling systems.
The photon energy ($\sim$ 2 eV) typically used in experiments is much larger than the energy scale of the low-energy model. Accurate calculation of the Raman intensity requires detailed electron bands far from the Fermi surface, which is extremely challenging.
Therefore, we primarily focus on the peak positions of the Raman spectrum and its pressure dependence, which can be accurately calculated using our approach.

To efficiently calculate the resonant part away from the strong coupling limit, we derive a mean-field Raman operator for low-energy excitations. Assuming the Hamiltonian under mean field approximation is expressed as
\begin{equation}
\hat{\mathcal{H}}_{\rm mf}= 
\sum_{\boldsymbol{k} m } (\epsilon_{m \bm{k}} - \mu) \psi_m^{\dagger}(\boldsymbol{k})  \psi_m(\boldsymbol{k}) - E_0,
\end{equation}
where $E_0$ is the mean field ground state energy, $\mu$ is the chemical potential, and $\psi_m(\boldsymbol{k})$ is the Bloch wave function. 
Note that $\hat{\mathcal{H}}_{\rm mf} |0\rangle =0$, where $|0\rangle$ is the ground state of the mean field Hamiltonian.
The mean-field Hamiltonian satisfies the commutation relation:
\begin{eqnarray}
[\hat{\mathcal{H}}_{\rm mf}, \psi_m^{\dagger}(\boldsymbol{k}) \psi_n(\boldsymbol{k})]  &=& (\varepsilon_{m\bm{k}} - \varepsilon_{n \bm{k}} ) \psi_m^{\dagger}(\boldsymbol{k}) \psi_n(\boldsymbol{k}),
\end{eqnarray}
which leads to 
\begin{equation}    
\hat{\mathcal{H}}_{\rm mf} \psi_m^{\dagger}(\boldsymbol{k}) \psi_n(\boldsymbol{k}) |0\rangle = (\varepsilon_{m\bm{k}} - \varepsilon_{n \bm{k}} ) \psi_m^{\dagger}(\boldsymbol{k}) \psi_n(\boldsymbol{k}) |0\rangle.
\end{equation}
Similarly,
\begin{eqnarray}
\hat{\mathcal{H}}^{p}_{\rm mf} \psi_m^{\dagger}(\boldsymbol{k}) \psi_n(\boldsymbol{k}) |0\rangle =
(\varepsilon_{m\bm{k}} - \varepsilon_{n \bm{k}} )^{p} \psi_m^{\dagger}(\boldsymbol{k}) \psi_n(\boldsymbol{k})) |0\rangle. \ \ 
\end{eqnarray}
for any integer $p \geq 0$. 
Thus, for an analytic function $f$,
\begin{equation}
f(\hat{\mathcal{H}}_{\rm mf}) \psi_m^{\dagger}(\boldsymbol{k}) \psi_n(\boldsymbol{k}) |0\rangle = f(\varepsilon_{m\bm{k}} - \varepsilon_{n \bm{k}})  \psi_m^{\dagger}(\boldsymbol{k}) \psi_n(\boldsymbol{k}) |0\rangle.
\end{equation}

A \emph{mean field} Raman operator is obtained by approximating the exact resolvent in Eq.~\eqref{eq:M_R} with its mean field expression $\hat{ \bar{G}}_E = (\hat{\mathcal{H}}_{\rm mf} -E)^{-1} $.
As we will see below, this significant simplification is still sufficient to capture the contribution of the longitudinal mode to the low-frequency Raman cross section and the mean Raman operator reduces to the effective Raman operator derived in the large-$U$ limit~\cite{ShastryS1990,ShastryS1991}.
From the commutation relations, we obtain
\begin{equation}
\hat{ \bar{G}}_{\omega} \psi_m^{\dagger}(\boldsymbol{k}) \psi_n(\boldsymbol{k}) |0 \rangle =
\frac{1}{\varepsilon_{m\bm{k}} - \varepsilon_{n \bm{k}} - \omega}
\psi_m^{\dagger}(\boldsymbol{k}) \psi_n(\boldsymbol{k})  |0 \rangle.
\end{equation}
The $\mu$-component of the current can be expressed as
\begin{equation}
j_{\mu}(\bm{q}) =\sum_{\bm{k}mn}\psi_{m}^{\dagger}(\bm{k}-\bm{q}/2)v_{\mu}^{mn}(\bm{k},\bm{q})\psi_{n}(\bm{k}+\bm{q}/2),
\label{eq:j_mu}
\end{equation}
where $v_{\mu}^{mn}(\bm{k},\bm{q})$ is the velocity~\cite{suppl}.
Given the high speed of light, the momenta of photons can be approximately set to $\bm{k}_{i},\bm{k}_{f}=\bm{0}$.
Using these, the resonant part of the Raman operator~\eqref{eq:M_R} can be approximated by $\hat{M}_{R}^{\mu \nu}(\boldsymbol{q}) \approx \hat{\bar{M}}_{R_1}^{\mu \nu}(\boldsymbol{q}) + \hat{\bar{M}}_{R_2}^{\mu \nu}(\boldsymbol{q})$ with
\begin{widetext}
\begin{equation}
\begin{split}
\hat{\bar{M}}_{R_1}^{\mu \nu}(\boldsymbol{q}) &= \sum_{\boldsymbol{k} m n} \sum_{\boldsymbol{k}^{\prime} m^{\prime} n^{\prime}} \psi_{m^{\prime}}^{\dagger}(\boldsymbol{k}^{\prime}) v_\nu^{m^{\prime} n^{\prime}}(\boldsymbol{k}^{\prime}, \boldsymbol{k}_f) \psi_{n^{\prime}}(\boldsymbol{k}^{\prime}) \frac{1}{\varepsilon_{m\bm{k}} - \varepsilon_{n \bm{k}} - \omega_i} \psi_m^{\dagger}(\boldsymbol{k}) v_\mu^{m n}(\boldsymbol{k},-\boldsymbol{k}_i) \psi_n(\boldsymbol{k}),\\
\hat{\bar{M}}_{R_2}^{\mu \nu}(\boldsymbol{q}) &= \sum_{\boldsymbol{k}^{\prime} m^{\prime} n^{\prime}} \sum_{\boldsymbol{k} m n} \psi_m^{\dagger}(\boldsymbol{k}) v_\mu^{m n}(\boldsymbol{k},-\boldsymbol{k}_i) \psi_n(\boldsymbol{k}) 
\frac{1}{\varepsilon_{m'\bm{k}'} - \varepsilon_{n' \bm{k}'} + \omega_f} 
\psi_{m^{\prime}}^{\dagger}(\boldsymbol{k}^{\prime}) v_\nu^{m^{\prime} n^{\prime}}(\boldsymbol{k}^{\prime}, \boldsymbol{k}_f) \psi_{n^{\prime}}(\boldsymbol{k}^{\prime}),
\end{split}
\label{eq:M_R_mf}
\end{equation}
\end{widetext}
when they act on the ground state at zero temperature~\cite{suppl}.
In the strong coupling case, the energy differences $\varepsilon_{m\bm{k}} - \varepsilon_{n \bm{k}}$ and  $\varepsilon_{m'\bm{k}'} - \varepsilon_{n' \bm{k}'}$ in the denominators of Eq.~\eqref{eq:M_R_mf} are well approximated by $U$, implying that that the mean field resolvent $\hat{ \bar{G}}_{\omega}$ produces the same denominator, $1/(U-\omega)$ with $\omega=\omega_i$ and $\omega=-\omega_f$,
as found in degenerate perturbation theory of the exact Hamiltonian. In other words, at low enegies the
the mean field Raman operator $\hat{\bar{M}}_{R_{1,2}}^{\mu \nu}(\boldsymbol{q})$ reduces to the  effective spin operator as in the Fleury-Loudon theory~\cite{ShastryS1990,ShastryS1991,Fleury68,Elliott63}.
Our approach is thus a simple extension of the established strong coupling approach to the intermediate coupling regime.

\begin{figure}[tb!]
    \centering
    \includegraphics[width=0.95\columnwidth]{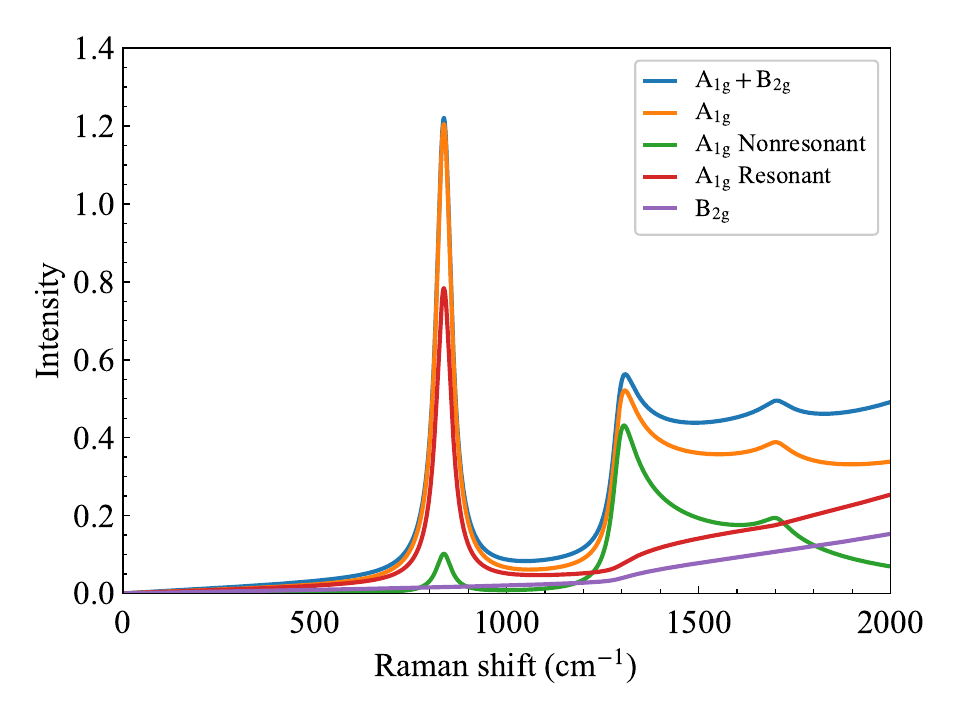}
    \caption{Raman spectrum for the $A_{1g}$ and $B_{2g}$ channels calculated using the effective operator for single particle-hole pair excitation. The $A_{1g}$ channel has both resonant and nonresonant parts, while the $B_{2g}$ channel only has the resonant contribution. The model parameters are optimized for Sr$_3$Ir$_2$O$_7$, reproducing the lowest ($\approx850$ cm$^{-1}$) and second lowest ($\approx1300$ cm$^{-1}$) energy peak positions.}
    \label{fig:Raman}
\end{figure}

\textit{Results.}---We use the Hartree-Fock (mean field) approximation to determine the magnetization $M$ and calculate the Raman spectrum of the bilayer Hubbard model using the effective operator and the random phase approximation (RPA)~\cite{suppl}. 
According to Eq.~\eqref{eq:cross-section}, the Raman scattering cross-section is proportional to the Fourier transform of the two-point correlation function:
\begin{equation}
R^{\bm{e}^i \bm{e}^f}(\omega) = \int_{-\infty}^{\infty} dte^{i\omega t}\langle {\bm e}^i_{\mu} {\bm e}^f_{\nu} \hat{\bar{M}}^{\mu \nu \dagger} (t)  {\bm e}^i_{\alpha} {\bm e}^f_{\beta} \hat{\bar{M}}^{\alpha \beta}(0)\rangle.
\end{equation}
Although the resonant part includes contributions from two particle-hole pair states, single particle-hole pair states dominate the low-energy excitations near the quantum phase transition (QPT) between the excitonic condensate and the paramagnetic state. 
Therefore, in this study, we focus exclusively on single particle-hole pair states~\cite{suppl}.
Figure~\ref{fig:Raman} shows the calculated Raman spectrum.
The model parameters are optimized for Sr$_3$Ir$_2$O$_7$ to match the experimentally observed lowest ($\approx850$ cm$^{-1}$) and second lowest ($\approx1300$ cm$^{-1}$) energy peak positions, while maintaining consistency with the RIXS data~\cite{Kim2012_2,Moretti2015evidence,Mazzone2022} and the density functional theory calculations~\cite{Moon2008,Wang2013}. The optimized Hubbard Hamiltonian parameters are $t_1 = 0.12$, $t_z = 0.092$, $U = 0.33$ eV, $t_2/t_1 = 0.1$, and $\alpha = 1.45$~\cite{Suwa2021,Mazzone2022}.
We chose $\omega_i=0.5$ eV to get divergent resonant contributions within the low-energy model. In calculations, we shift the imaginary parts of the frequencies, $\omega_i$ and $\omega_f$. The imaginary part determines the peak width, optimized to $3\times 10^{-3}$ to match the experimental observation in the present results.
Our main conclusions, including the peak positions and symmetry, remain unaffected by the choice of incident photon energy.

\begin{figure}[bt]
    \centering
    \includegraphics[width=\columnwidth]{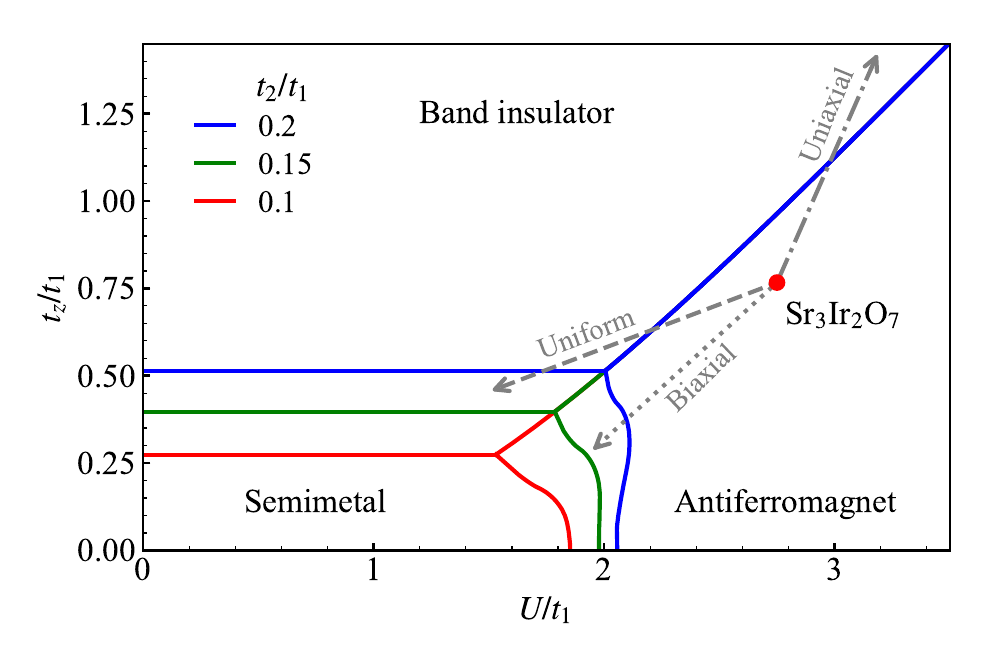}
    \caption{Phase diagram of the bilayer Hubbard model for $t_2/t_1=0.1,0.15$, and $0.2$. The transitions to the band insulator are continuous, while the transition between the semimetal and the antiferromagnetic phases is first order. The parameter for Sr$_3$Ir$_2$O$_7$ at ambient pressure is indicated by the red circle. The expected paths under uniform, uniaxial, and biaxial pressure are shown by the dashed, dash-dot, and dotted lines, respectively.
    In the uniaxial (biaxial) case, the pressure is assumed to be applied on the $z$ ($xy$) axis.
    The boundary between the band insulator and the antiferromagnet for $t_2/t_1=0.2$ overlaps with the boundaries for $t_2/t_1=0.1$ and $0.15$.
    }
    \label{fig:pd}
\end{figure}
\begin{figure*}[ht]
    \centering
    \includegraphics[width=\textwidth]{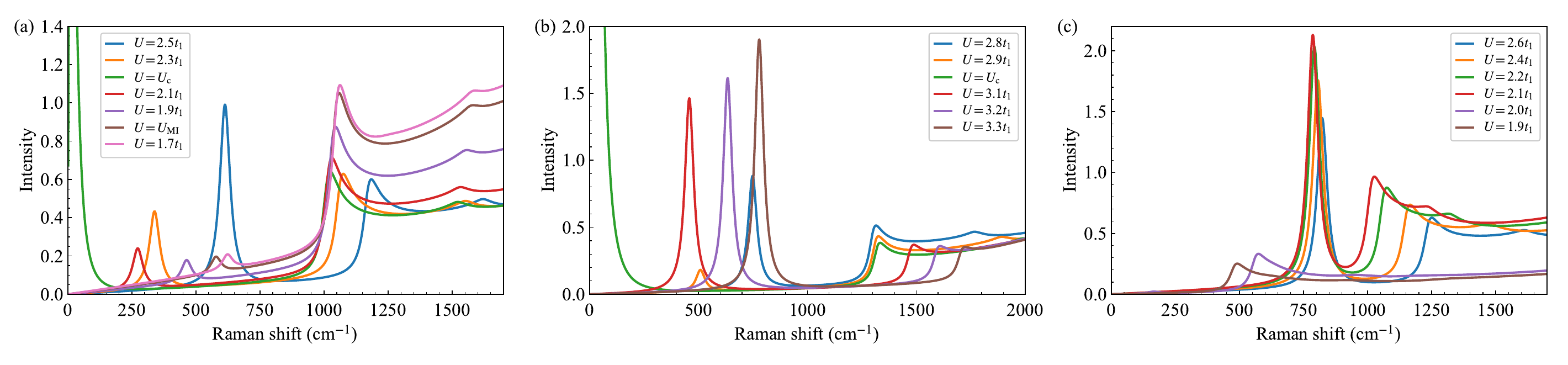}
    \caption{Raman spectra for several values of $U$ under pressure. (a) Uniform (hydrostatic) pressure: The longitudinal mode shows significant softening, and the exciton mode in the band-insulator phase beyond $U_{\rm c}$ has a small intensity. 
    No significant change appears at the semimetal-insulator transition ($U_{\rm MI}$).
    (b) Uniaxial ($z$-axis) pressure: The exciton-mode intensity is enhanced in the band-insulator phase in comparison to that in the uniform case. (c) Biaxial ($xy$-axes) pressure: No softening occurs at the first-order transition to the semimetallic phase.}
    \label{fig:Raman-2}
\end{figure*}

A significant finding of the present work is that the lowest energy peak of the Raman spectrum is not a transverse (magnon) mode but the longitudinal mode, which is active in the $A_{1g}$ channel ($(R^{xx}+R^{yy})/2$~\footnote{The $B_{1g}$ channel has no contribution due to the $C_4$ symmetry.}) but inactive in the $B_{2g}$ channel ($R^{xy}$), in agreement with experimental observations~\cite{Li2020symmetry}.
The second-lowest energy peak arises from the peak density of states in the particle-hole continuum, followed by a long tail due to its continuum nature. Notably, the lowest energy peak is dominated by resonant contributions, while the second-lowest energy peak is dominated by non-resonant contributions.

An additional peak was experimentally observed at $1400$ $\rm{cm}^{-1} \sim 170$ $\rm{meV}$~\cite{Gretarsson2016two,Zhang2019,Li2020symmetry}. This excitation is most likely arising from the two-magnon continuum, as the excitation energy matches twice the magnon gap of $85$ $\rm{meV}$ measured in the RIXS~\cite{Moretti2015evidence,Mazzone2022}. Although our calculations do not include two-magnon excitations formed by two particle-hole pairs, they accurately reproduce the lowest and second-lowest energy peaks of the Raman spectrum. Importantly, the lowest energy peak arises from the longitudinal mode that becomes soft  at the QPT between the excitonic condensate and the paramagnetic state.

To understand the measured pressure dependence~\cite{Li2013,Zhao2014,Zocco2014,Ding2016,Donnerer2018}, we obtain the phase diagram with respect to $U/t_1$ and $t_z/t_1$ in the mean field approximation, as shown in Fig.~\ref{fig:pd}, fixing the other Hamiltonian parameters.
Under hydrostatic (uniform) pressure, the lattice constant within each layer decreases more than between layers~\cite{Zhao2014}, causing $t_z/t_1$ to decrease along with $U/t_1$. 
For the expected path in the phase diagram under pressure, shown in Fig.~\ref{fig:pd}, the ratio $t_2/t_1$ is expected to increase because the angle of the staggered octahedral rotation in the plane increases~\cite{Ding2016}.
Although the ratio $t_2/t_1$ is estimated to be $0.1$ from RIXS measurements at ambient pressure, it can become significantly larger under uniform (and also biaxial) pressure.
Thus, as shown in Fig.~\ref{fig:Raman-2}, after setting $t_2/t_1=0.2$~\footnote{In our approximation, using a different $t_2$ only shifts the phase boundary without changing the Raman spectrum within the same phase.}, we calculate the Raman spectra for several values of $U$ on the path.
As indicated by the softening of the longitudinal mode (Fig.~\ref{fig:Raman-2}(a)), the antiferromagnetic system initially transitions to the band-insulator phase for uniform pressure, as observed in experiments~\cite{Zhang2019}. 
The longitudinal mode transforms into an exciton mode in the band-insulator phase, whose intensity is small and may not be clearly observed in experiments.
The transverse (magnon) mode does not exhibit any softening during this transition due to the easy $z$-axis spin anisotropy.
As pressure increases, the system subsequently transitions to the semimetallic phase, resulting in the closure of the indirect band gap.
These findings are overall consistent with the experimental results~\cite{Zhang2019,Ding2016}. Although the value of $\alpha$ might change slightly under pressure, it does not significantly affect the phase diagram~\cite{Suwa2021}.

We also study the effect of pressure anisotropy and show the unique controllability of the BCS-BEC crossover. 
Under uniaxial pressure on the $z$ axis, $t_z$ should increase and $t_1$ decrease as the system shrinks along the $z$ axis while keeping the total volume. Under biaxial pressure on the $xy$ axes~\footnote{Experimental realizations of biaxial pressure have been discussed and implemented for several materials~\cite{Nakajima2021}.}, $t_z$ should decrease and $t_1$ increase. 
The expected paths are also shown in Fig.~\ref{fig:pd}.
In the uniaxial case, the longitudinal mode shows significant softening similar to the uniform case, and the excitonic mode in the band-insulator phase has enhanced intensity (Fig.~\ref{fig:Raman-2}(b)). 
The exciton condensation from the disordered phase can be clearly observed under uniaxial pressure as strong evidence of the BEC scenario.
In the biaxial case, no softening occurs at the first-order transition to the semimetallic phase (Fig.~\ref{fig:Raman-2}(c)). This indicates that exciton formation and condensation occur simultaneously at the transition point, which is consistent with the BCS scenario~\cite{Nozieres1985,Ohashi2020}.

\textit{Conclusions.}---
We provide a theoretical account of the magnetic Raman spectrum of the bilayer iridate Sr$_3$Ir$_2$O$_7$. Our results reveal that the peak in the low-energy Raman spectrum identified as $M_1$~\cite{Li2020symmetry} corresponds to a longitudinal magnetic mode. Therefore, the observed softening of this longitudinal mode under uniform pressure~\cite{Zhang2019} provides direct evidence of the excitonic insulator character of the antiferromagnetic state. Indeed, this material is uniquely close to the QPT induced by triplet exciton condensation, and its BCS-BEC crossover can be controlled by isotropic and anisotropic pressure. To fully account for the asymmetric broadening of the longitudinal mode peak~\cite{Li2020symmetry}, it would be necessary to include exciton-phonon interactions~\cite{Antonius2022}. We leave this aspect for future research.

\begin{acknowledgments}
The authors are grateful to Liuyan Zhao and Mark P. M. Dean for fruitful discussions.
Calculations were performed using computational resources of the Supercomputer Center at the Institute for Solid State Physics, the University of Tokyo.
H.S. acknowledges Inamori Research Grants from Inamori Foundation and support from JSPS KAKENHI Grant Nos. JP22K03508 and JP24H01609. C.D.B. and S-S. Z. acknowledge support from the U.S.~Department of Energy, Office of Science, Office of Basic Energy Sciences, under Award Number DE-SC0022311. 
\end{acknowledgments}

\clearpage
\widetext

\begin{center}
  \textbf{
    \begin{large}
      Supplemental Material: Electronic Raman scattering of antiferromagnetic excitonic insulators      
    \end{large}
  }
\end{center}

\renewcommand{\thefigure}{S\arabic{figure}}
\setcounter{figure}{0}
\renewcommand{\theequation}{S\arabic{equation}}
\setcounter{equation}{0}
\renewcommand{\citenumfont}[1]{S#1}
\renewcommand{\bibnumfmt}[1]{[S#1]}

\section{Mean field approximation}
We model Sr$_3$Ir$_2$O$_7$ using a $J_{\rm eff}=1/2$ single-orbital bilayer square-lattice Hubbard Hamiltonian $\mathcal{H}=-\mathcal{H}_{{\rm K}}+\mathcal{H}_{{\rm I}}$ at half-filling,
with $\mathcal{H}_{{\rm I}}\!=U\sum_{\bm{r}}n_{\bm{r}{\uparrow}}n_{\bm{r}{\downarrow}}$
and 
\begin{eqnarray}
\!\!\!\mathcal{H}_{{\rm K}}\!=\sum_{{\bm{r}},\bm{\delta}_{a}}t_{a}{\bm{c}}_{\bm{r}}^{\dagger}{\bm{c}}_{{\bm{r}}+{\bm{\delta}}_{a}}\!\!+t_{z}\!\!\sum_{{\bm{r}}_{\bot}}\!{\bm{c}}_{({\bm{r}}_{\bot},1)}^{\dagger}e^{i\frac{\alpha}{2}\epsilon_{\bm{r}}\sigma_{z}}{\bm{c}}_{({\bm{r}}_{\bot},2)}\!\!+\!{\rm H.c.},\label{eq:H_bl}
\end{eqnarray}
where 
$t_a \in\mathbb{R}$ ($a$ = 1, 2)
are the nearest- and next-nearest-neighbor hopping amplitudes within the square lattice of each Ir layer, 
$t_z \in\mathbb{R}$ is the hopping amplitude on the vertical bonds between the two layers, and $\sigma_z$ is the Pauli matrix. 
The overall phase of the orbital states (i.e., Wannier wave functions) was chosen to gauge away the phase of 
$t_a$.
The operator $ c_{\bm{r}}^{\dagger}$ = [$ c_{\bm{r}\uparrow}^{\dagger}$, $c_{\bm{r}\downarrow}^{\dagger}$] creates the 
two-component spinor of the electron field at $\bm{r} = (\bm{r}_\bot, l)$ with $l = 1, 2$ denoting the layer index and $\bm{r}_\bot = r_1\bm{a}_1 + r_2\bm{a}_2$ ($r_1,r_2 \in {\mathbb{Z}}$), with $\bm{a}_1$ and $\bm{a}_2$ being the primitive in-plane lattice vectors. The directed nearest and next-nearest neighboring bonds are represented by $\bm{\delta}_1=\bm{a}_1,\bm{a}_2$ and $\bm{\delta}_2=\bm{a}_1 \pm \bm{a}_2$. 
The sign $\epsilon_{\bm{r}}$ of the spin-dependent hopping term takes the values $\pm1$ depending on which sublattice ($\mathcal{A}$ or $\mathcal{B}$) of the bipartite bilayer system $\bm{r}$ points to, where ${\cal A}$ and ${\cal B}$ correspond to the case where ${\rm mod}(r_1+r_2+l,2)=0$. 
The key phase $\alpha$ arises from hopping matrix elements between $d_{xz}$ and $d_{yz}$ orbitals, which are allowed to be nonzero by the staggered octahedral rotation in the unit cell~\cite{Carter2013microscopicS}.
A finite $\alpha\neq0$ gives rise to an easy
$z$-axis spin anisotropy, for which case, the putative N\'eel ordering would be along this easy axis, $\langle S_{\bm{r}}^{\mu}\rangle=(-1)^{\gamma_{\bm{r}}}M\delta_{\mu z}$,
where $\gamma_{\bm{r}}=(1+\epsilon_{\bm{r}})/2$, ${S}_{\bm{r}}^{\mu}={1/2}c_{\bm{r}}^{\dagger}\sigma^{\mu}c_{\bm{r}}$
$(\mu=x,y,z)$ and $M$ is the magnetization. In the interaction term $H_{\rm I}$, $U$ is the effective on-site Coulomb interaction, and $n_{\bm{r}\sigma}$ is the number operator of spin-$\sigma$ electrons at $\bm{r}$.

Following Ref.~\cite{Suwa2021S},
a mean-field decoupling of $\mathcal{H}_{{\rm I}}$ leads to ${\cal H}^{{\rm MF}}=\sum_{\bm{k}\in{\rm BZ}}{\bm{c}}_{\bm{k}}^{\dagger}{\cal H}_{\bm{k}}^{{\rm MF}}{\bm{c}}_{\bm{k}}^{\;}$
with 
\begin{align}
{\cal H}_{\bm{k}}^{{\rm MF}} & =\left(\begin{array}{cc}
\epsilon_{\bm{k}}^{(2)}+UM\sigma_{z} & \epsilon_{\bm{k}}^{(1)}-t_{z}\cos(k_{z})e^{-i\frac{\alpha}{2}\sigma_{z}}\\
\epsilon_{\bm{k}}^{(1)}-t_{z}\cos(k_{z})e^{i\frac{\alpha}{2}\sigma_{z}} & \epsilon_{\bm{k}}^{(2)}-UM\sigma_{z}
\end{array}\right),
\end{align}
where ${\bm{c}}_{\bm{k}}\equiv(c_{\mathcal{A}\uparrow,\bm{k}},c_{\mathcal{A}\downarrow,\bm{k}},c_{\mathcal{B}\uparrow,\bm{k}},c_{\mathcal{B}\downarrow,\bm{k}})^{T}$
is the Fourier-transformed operator, 
\begin{align}
\epsilon_{\bm{k}}^{(1)} & =-2t_{1}\left(\cos\frac{k_{1}+k_{2}}{2}+\cos\frac{k_{1}-k_{2}}{2}\right),\\
\epsilon_{\bm{k}}^{(2)} & =-2t_{2}\left(\cos k_{1}+\cos k_{2}\right),
\end{align}
and ${\bm{k}}\equiv({\bm{k}}_{\bot},k_{z})$ a wavevector in the first Brillouin
zone (BZ) \emph{i.e.} ${\bm{k}}_{\bot}=k_{1}{\bm{b}}_{1}^{\prime}+k_{2}{\bm{b}}_{2}^{\prime}$,
with ${\bm{b}}_{1}^{\prime}=(1/2,-1/2)$, ${\bm{b}}_{2}^{\prime}=(1/2,1/2)$,
{$k_{1},k_{2}\in[0,2\pi)$} and $k_{z}=0,\pi$. ${\cal H}_{\bm{k}}^{{\rm MF}}$ is diagonalized by a $4\times4$
unitary matrix $U(\bm{k})$, yielding 
\begin{eqnarray}
c_{\gamma\sigma,\bm{k}} & = & \sum_{n}U_{(\gamma\sigma),n}(\bm{k})\psi_{n,\bm{k}},
\end{eqnarray}
where $n\equiv(s,\sigma)$ with $s=\pm$, $\sigma=\uparrow,\downarrow$, $\psi_{n,\bm{k}}$ is the Bloch wave function,
and each column of $U(\bm{k})$ is an eigenvector of ${\cal H}_{\bm{k}}^{{\rm MF}}$:
\begin{eqnarray}
X_{s\uparrow}(\bm{k})\!=\!\left(\begin{array}{c}
x_{\uparrow\bm{k}}\sqrt{\frac{1+sz_{\uparrow\bm{k}}}{2}}\\
0\\
s\sqrt{\frac{1-sz_{\uparrow\bm{k}}}{2}}\\
0
\end{array}\right),\;X_{s\downarrow}(\bm{k})\!=\!\left(\begin{array}{c}
0\\
x_{\downarrow\bm{k}}\sqrt{\frac{1-sz_{\downarrow\bm{k}}}{2}}\\
0\\
s\sqrt{\frac{1+sz_{\downarrow\bm{k}}}{2}}
\end{array}\right),\quad\ 
\label{eq:eigenvec}
\end{eqnarray}
where 
\begin{eqnarray}
x_{\sigma\bm{k}} & = & \frac{b_{\sigma\bm{k}}}{\rvert b_{\sigma\bm{k}}\rvert},\;\;z_{\sigma\bm{k}}=\frac{\delta}{\sqrt{\delta^{2}+\rvert b_{\sigma\bm{k}}\rvert^{2}}},\;
b_{\sigma\bm{k}}= \epsilon_{\bm{k}}^{(1)}-t_{z}\cos(k_{z})e^{-i\sigma\frac{\alpha}{2}}
\end{eqnarray}
and $\delta=UM$ determined by solving the self-consistent equation \cite{Suwa2021S}. The corresponding eigenenergy, $\varepsilon_{s\sigma}(\bm{k})=\epsilon_{\bm{k}}^{(2)}+s\sqrt{\delta^{2}+\rvert b_{\sigma\bm{k}}\rvert^{2}},$
is independent of the spin flavour, $\rvert b_{\sigma\bm{k}}\rvert^{2}=b_{\bm{k}}^{2}$
and $z_{\sigma\bm{k}}\equiv z_{\bm{k}}$.

\section{Electronic Raman scattering cross section}
The interaction between electrons
and light can be treated by using  the Peierls construction.
For simplicity, we temporarily switch to a generic notation, where $j$ denotes the lattice site, 
$c_j\equiv(c_{j\uparrow},c_{j\downarrow})^T$ denotes the two-component fermionic operator at site $j$, 
and $t_{j\ell \alpha \beta}$ denotes the hopping amplitude between sites $j$ and $\ell$, where $\alpha,\beta=\uparrow,\downarrow$ refer to the spin component.
The resulting
kinetic term is expressed by
\begin{equation}
- H_{{\rm K}}'=\sum_{j\ell\alpha\beta}t_{j\ell\alpha\beta}c_{j\alpha}^{\dagger}c_{\ell\beta}e^{i\frac{e}{\hbar c}\int_{\ell}^{j}\bm{A}\cdot d\bm{s}}.
\end{equation}
Assuming that the spatial variation of the vector potential is much slower than the lattice spacing scale, the integral of the vector potential can be approximated by
\begin{equation}
\int_{\bm{r}}^{\bm{r}+\bm{x}}\bm{A}\cdot d\bm{s}\approx\bm{A}(\bm{r}+\frac{\bm{x}}{2})\cdot\bm{x}.
\end{equation}
Expanding in powers of $\bm{A}$ up to the second order leads to the
scattering Hamiltonian \cite{ShastryS1990S,ShastryS1991S}
\begin{equation}
H_{{\rm int}}=-\frac{e}{\hbar c}\sum_{\bm{q}}\bm{j}(\bm{q})\cdot\bm{A}(-\bm{q})+\frac{e^{2}}{2\hbar^{2}c^{2}}\sum_{\bm{q}_{1}\bm{q}_{2}\mu\nu}A_{\mu}(-\bm{q}_{1})\tau_{\mu\nu}(\bm{q}_{1}+\bm{q}_{2})A_{\nu}(-\bm{q}_{2}),
\end{equation}
where $\bm{j}(\bm{q})$ is the Fourier transform of the current density operator, $\tau$ is the second rank stress tensor, and we consider the in-plane components $\mu,\nu=x,y$. 
The cross section
is determined by the transition probability rate $R$ obtained from Fermi's golden rule:
\begin{align}
R^{\bm{e}^i \bm{e}^f} \!\!(\bm{q},\Omega)
= 2\pi g(\bm{k}_{i})^2 g(\bm{k}_{f})^2 \sum_{\eta\gamma \mu \nu} e^{-\beta\epsilon_{\eta}} \rvert {\bm e}^i_{\mu} {\bm e}^f_{\nu} \langle \gamma \rvert \hat{M}^{\mu\nu}(\bm{q})\rvert \eta\rangle\rvert^{2}
\delta(\epsilon_{\gamma}-\epsilon_{\eta}-\Omega)
\label{eq:cross-section2}
\end{align}
where $\rvert \eta\rangle$ $(\rvert \gamma \rangle)$ is the initial (final)
eigenstate of the system with energy $\epsilon_{\eta(\gamma)}$, $\beta$ the inverse of temperature,  
$\omega_{i(f)}$, $\bm{k}_{i(f)}$, and $\bm{e}^{i(f)}$ the energy, momentum, and polarization of the initial (final) state of a photon, respectively, $\bm{q}=\bm{k}_{i}-\bm{k}_{f}$, $\Omega=\omega_{i}-\omega_{f}$ the momentum and frequency transfer during the two-Raman scattering process, and $g(\bm{q})=(hc^{2}/V\omega_{\bm{q}})^{1/2}$ the electron-photon coupling constant, where $V$ is the volume.
The scattering operator reads $\hat{M}^{\mu\nu}(\bm{q})=\hat{M}^{\mu\nu}_{N}(\bm{q})+\hat{M}^{\mu\nu}_{R}(\bm{q})$
which includes a nonresonant contribution
\begin{equation}
\langle \gamma \rvert \hat{M}_{N}^{\mu\nu}(\bm{q})\rvert \eta\rangle=\langle \gamma \rvert\tau_{\mu\nu}(\bm{q})\rvert \eta\rangle,
\end{equation}
and a resonant contribution
\begin{equation}
\langle \gamma \rvert \hat{M}_{R}^{\mu\nu}(\bm{q})\rvert \eta\rangle=\sum_{\zeta}\frac{\langle \gamma \rvert j_{\nu}(\bm{k}_{f})\rvert\zeta\rangle\langle\zeta\rvert j_{\mu}(-\bm{k}_{i})\rvert \eta \rangle}{\epsilon_{\zeta}-\epsilon_{\eta}-\omega_{i}}+\frac{\langle \gamma \rvert j_{\mu}(-\bm{k}_{i})\rvert\zeta\rangle\langle\zeta\rvert j_{\nu}(\bm{k}_{f})\rvert \eta\rangle}{\epsilon_{\zeta}-\epsilon_{\eta}+\omega_{f}},
\end{equation}
where $j_{\mu}(\bm{k})=\bm{j}(\bm{k})\cdot\bm{e}_{\mu}$, and $\bm{e}_{\mu}$ is the unit vector in $\mu$ direction.  
Given the high speed of light and that $\omega_{i/f}$ is on the same energy scale as the electronic states, the momenta of photons can be safely set to $\bm{k}_{i},\bm{k}_{f}=\bm{0}$, and thus $\bm{q}=\bm{0}$.

In our model~\eqref{eq:H_bl}, the scattering operator for the nonresonant contribution is given by
\begin{align}
    \hat{M}^{\mu\nu}_N(\bm{q}) &= \sum_{\bm{k}\sigma} \left[ \sum_{\bm{\delta}_1} t_1 \left(e^{i\bm{k} \cdot \bm{\delta}_1} + e^{-i \bm{k} \cdot \bm{\delta}_1} \right) c^\dagger _{\mathcal{A}\sigma}(\bm{k} - \bm{q}/2) c_{\mathcal{B}\sigma}(\bm{k} + \bm{q}/2) \left( \bm{\delta}_1 \cdot \bm{e}_\mu \right)\left( \bm{\delta}_1 \cdot \bm{e}_\nu \right) \right.\nonumber \\
    & + \sum_{\bm{\delta}_1} \bar{t}_1 \left( e^{i\bm{k} \cdot \bm{\delta}_1} + e^{-i \bm{k} \cdot \bm{\delta}_1} \right) c^\dagger _{\mathcal{B}\sigma}(\bm{k} - \bm{q}/2) c_{\mathcal{A}\sigma}(\bm{k} + \bm{q}/2) \left( \bm{\delta}_1 \cdot \bm{e}_\mu \right)\left( \bm{\delta}_1 \cdot \bm{e}_\nu \right)   \label{eq:M_N} \\
    & + \left. \sum_{\bm{\delta}_2} \left(t_2 e^{i\bm{k} \cdot \bm{\delta}_2} + \bar{t}_2 e^{-i \bm{k} \cdot \bm{\delta}_2} \right) \left( c^\dagger _{\mathcal{A}\sigma}(\bm{k} - \bm{q}/2) c_{\mathcal{A}\sigma}(\bm{k} + \bm{q}/2) + c^\dagger _{\mathcal{B}\sigma}(\bm{k} - \bm{q}/2) c_{\mathcal{B}\sigma}(\bm{k} + \bm{q}/2) \right) \left( \bm{\delta}_2 \cdot \bm{e}_\mu \right)\left( \bm{\delta}_2 \cdot \bm{e}_\nu \right) \right] \nonumber.
\end{align}
In the case of $\bm{q}=\bm{0}$, it can be represented by
\begin{align}
        \hat{M}^{\mu\nu}_N(\bm{0}) = \sum_{\bm{k}} \bm{c}^\dagger_{\bm{k}} M^{\mu\nu}_{N,\bm{k}} \bm{c}_{\bm{k}},
        \label{eq:M_N0}
\end{align}
where $M^{\mu\nu}_{N,\bm{k}}$ is the corresponding matrix, as defined in Eq.~\eqref{eq:M_N}.
For the resonant contribution, the current operator is given by
\begin{align}
    \bm{j}(\bm{q}) &= -i\sum_{\bm{k}\sigma} \left[ \sum_{\bm{\delta}_1} \bm{\delta}_1 t_1 \left(e^{i\bm{k} \cdot \bm{\delta}_1} - e^{-i \bm{k} \cdot \bm{\delta}_1} \right) c^\dagger _{\mathcal{A}\sigma}(\bm{k} - \bm{q}/2) c_{\mathcal{B}\sigma}(\bm{k} + \bm{q}/2) \right.\nonumber \\
    & \quad + \sum_{\bm{\delta}_1} \bm{\delta}_1 \bar{t}_1 \left(e^{i\bm{k} \cdot \bm{\delta}_1} - e^{-i \bm{k} \cdot \bm{\delta}_1} \right) c^\dagger _{\mathcal{B}\sigma}(\bm{k} - \bm{q}/2) c_{\mathcal{A}\sigma}(\bm{k} + \bm{q}/2) \\
    & \quad + \left. \sum_{\bm{\delta}_2} \bm{\delta}_2 \left(t_2 e^{i\bm{k} \cdot \bm{\delta}_2} - \bar{t}_2 e^{-i \bm{k} \cdot \bm{\delta}_2} \right) \left( c^\dagger _{\mathcal{A}\sigma}(\bm{k} - \bm{q}/2) c_{\mathcal{A}\sigma}(\bm{k} + \bm{q}/2) + c^\dagger _{\mathcal{B}\sigma}(\bm{k} - \bm{q}/2) c_{\mathcal{B}\sigma}(\bm{k} + \bm{q}/2) \right) \right] \nonumber\\
        &\equiv \sum_{\bm{k}} \sum_{mn} \psi^\dagger_m (\bm{k} - \bm{q}/2) \bm{v}^{mn}_{xy}(\bm{k},\bm{q}) \psi_n
    (\bm{k} + \bm{q}/2),
    \label{eq:J}
\end{align}
where
\begin{align}
    \bm{v}^{mn}_{xy}(\bm{k},\bm{q})&=-i\sum_{\sigma} \left[ \sum_{\bm{\delta}_1} \bm{\delta}_1 t_1 \left(e^{i\bm{k} \cdot \bm{\delta}_1} - e^{-i \bm{k} \cdot \bm{\delta}_1} \right) U^\dagger _{m,\mathcal{A}\sigma}(\bm{k} - \bm{q}/2) U_{\mathcal{B}\sigma,n}(\bm{k} + \bm{q}/2) \right.\nonumber \\
    & \quad + \sum_{\bm{\delta}_1} \bm{\delta}_1 \bar{t}_1 \left(e^{i\bm{k} \cdot \bm{\delta}_1} - e^{-i \bm{k} \cdot \bm{\delta}_1} \right) U^\dagger _{m,\mathcal{B}\sigma}(\bm{k} - \bm{q}/2) U_{\mathcal{A}\sigma,n}(\bm{k} + \bm{q}/2) \\
    & \quad + \left. \sum_{\bm{\delta}_2} \bm{\delta}_2 \left(t_2 e^{i\bm{k} \cdot \bm{\delta}_2} - \bar{t}_2 e^{-i \bm{k} \cdot \bm{\delta}_2} \right) \left( U^\dagger _{m,\mathcal{A}\sigma}(\bm{k} - \bm{q}/2) U_{\mathcal{A}\sigma,n}(\bm{k} + \bm{q}/2) + U^\dagger _{m,\mathcal{B}\sigma}(\bm{k} - \bm{q}/2) U_{\mathcal{B}\sigma,n}(\bm{k} + \bm{q}/2) \right) \right] \nonumber.
\end{align}

\section{Effective Raman operator}
We derive here a mean field Raman operator for the resonant contribution to the low-energy excitation.
Let us consider $\mu$ component in the $xy$ plane. 
From Eqs.(\ref{eq:J}),
the $\mu$ component of the current is represented by 
\begin{align}
j_{\mu}(\bm{q}) & =\sum_{\bm{k}mn}\psi_{m}^{\dagger}(\bm{k})v_{\mu}^{mn}(\bm{k},\bm{q})\psi_{n}(\bm{k}),
\end{align}
where $v_{\mu}^{mn}(\bm{k},\bm{q})=\bm{v}_{xy}^{mn}(\bm{k},\bm{q})\cdot\bm{e}_{\mu}$.
The resonant contribution is expressed by
\begin{align}
& \left\langle \gamma \left| \hat{M}_R^{\mu \nu}(\boldsymbol{q})\right| \eta\right\rangle=\sum_\zeta\left(R_1+R_2\right), \\
R_1= & \frac{\left\langle \gamma \left|\sum_{\boldsymbol{k}^{\prime} m^{\prime} n^{\prime}} \psi_{m^{\prime}}^{\dagger}\left(\boldsymbol{k}^{\prime}\right) v_\nu^{m^{\prime} n^{\prime}}\left(\boldsymbol{k}^{\prime}, \boldsymbol{k}_f\right) \psi_{n^{\prime}}\left(\boldsymbol{k}^{\prime}\right)\right| \zeta\right\rangle\left\langle\zeta\left|\sum_{\boldsymbol{k m n}} \psi_m^{\dagger}(\boldsymbol{k}) v_\mu^{m n}\left(\boldsymbol{k},-\boldsymbol{k}_i\right) \psi_n(\boldsymbol{k})\right| \eta\right\rangle}{\epsilon_\zeta-\epsilon_\eta-\omega_i}, \\
R_2= & \frac{\left\langle \gamma \left|\sum_{\boldsymbol{k} m n} \psi_m^{\dagger}(\boldsymbol{k}) v_\mu^{m n}\left(\boldsymbol{k},-\boldsymbol{k}_i\right) \psi_n(\boldsymbol{k})\right| \zeta\right\rangle\left\langle\zeta\left|\sum_{\boldsymbol{k}^{\prime} m^{\prime} n^{\prime}} \psi_{m^{\prime}}^{\dagger}\left(\boldsymbol{k}^{\prime}\right) v_\nu^{m^{\prime} n^{\prime}}\left(\boldsymbol{k}^{\prime}, \boldsymbol{k}_f\right) \psi_{n^{\prime}}\left(\boldsymbol{k}^{\prime}\right)\right| \eta\right\rangle}{\epsilon_\zeta-\epsilon_\eta+\omega_f} .
\end{align}
We introduce here the Raman operators:
\begin{eqnarray}
 \hat{M}_{R_1}^{\mu \nu}(\boldsymbol{q}) &=& \sum_{\zeta} \frac{\sum_{\boldsymbol{k}^{\prime} m^{\prime} n^{\prime}} \psi_{m^{\prime}}^{\dagger}\left(\boldsymbol{k}^{\prime}\right) v_\nu^{m^{\prime} n^{\prime}}\left(\boldsymbol{k}^{\prime}, \boldsymbol{k}_f\right) \psi_{n^{\prime}}\left(\boldsymbol{k}^{\prime}\right) | \zeta \rangle \langle \zeta |\sum_{\boldsymbol{k} m n} \psi_m^{\dagger}(\boldsymbol{k}) v_\mu^{m n}\left(\boldsymbol{k},-\boldsymbol{k}_i\right) \psi_n(\boldsymbol{k})}{\epsilon_\zeta-\epsilon_\eta-\omega_i}, \\
\hat{M}_{R_2}^{\mu \nu}(\boldsymbol{q}) &=&  \sum_{\zeta}  \frac{\sum_{\boldsymbol{k} m n} \psi_m^{\dagger}(\boldsymbol{k}) v_\mu^{m n}\left(\boldsymbol{k},-\boldsymbol{k}_i\right) \psi_n(\boldsymbol{k}) | \zeta \rangle \langle\zeta |\sum_{\boldsymbol{k}^{\prime} m^{\prime} n^{\prime}} \psi_{m^{\prime}}^{\dagger}\left(\boldsymbol{k}^{\prime}\right) v_\nu^{m^{\prime} n^{\prime}}\left(\boldsymbol{k}^{\prime}, \boldsymbol{k}_f\right) \psi_{n^{\prime}}\left(\boldsymbol{k}^{\prime}\right)}{\epsilon_\zeta-\epsilon_\eta+\omega_f}.
\end{eqnarray}
These equations can be rewritten in terms of the resolvent:
\begin{eqnarray}
 \hat{M}_{R_1}^{\mu \nu}(\boldsymbol{q}) &=&  \sum_{\boldsymbol{k}^{\prime} m^{\prime} n^{\prime}} \psi_{m^{\prime}}^{\dagger}\left(\boldsymbol{k}^{\prime}\right) v_\nu^{m^{\prime} n^{\prime}}\left(\boldsymbol{k}^{\prime}, \boldsymbol{k}_f\right) \psi_{n^{\prime}}\left(\boldsymbol{k}^{\prime}\right) \frac{1}{\hat{H} -\epsilon_\eta-\omega_i} \sum_{\boldsymbol{k} m n} \psi_m^{\dagger}(\boldsymbol{k}) v_\mu^{m n}\left(\boldsymbol{k},-\boldsymbol{k}_i\right) \psi_n(\boldsymbol{k}),\\
\hat{M}_{R_2}^{\mu \nu}(\boldsymbol{q}) &=&   \sum_{\boldsymbol{k} m n} \psi_m^{\dagger}(\boldsymbol{k}) v_\mu^{m n}\left(\boldsymbol{k},-\boldsymbol{k}_i\right) \psi_n(\boldsymbol{k}) \frac{1}{\hat{H} -\epsilon_\eta+\omega_f} \sum_{\boldsymbol{k}^{\prime} m^{\prime} n^{\prime}} \psi_{m^{\prime}}^{\dagger}\left(\boldsymbol{k}^{\prime}\right) v_\nu^{m^{\prime} n^{\prime}}\left(\boldsymbol{k}^{\prime}, \boldsymbol{k}_f\right) \psi_{n^{\prime}}\left(\boldsymbol{k}^{\prime}\right).
\end{eqnarray}

A \emph{mean field} Raman operator is obtained by replacing the exact resolvent by its mean field approximation:
\begin{equation}
\frac{1}{\hat{H} -\epsilon_\eta-\omega_i} \simeq \frac{1}{\hat{H}_{\rm MF} -\omega_i},
\end{equation}
where the mean field Hamiltonian is
\begin{equation}
\hat{H}_{\rm MF}= 
\sum_{\boldsymbol{k} m } (\varepsilon_{m \bm{k}} - \mu) \psi_m^{\dagger}(\boldsymbol{k})  \psi_m(\boldsymbol{k}) - E_0,
\label{MFHamil}
\end{equation}
$E_0$ is the mean field ground state energy of the half-filled system, and we have used that $\epsilon_i \simeq E_0$ is the ground state energy of $\hat{H}$. 
We note that this choice leads to
\begin{equation}
\hat{H}_{\rm MF} |0\rangle_{\rm MF} =0,
\end{equation}
where $|0\rangle_{\rm MF}$ is the ground state of the mean field Hamiltonian.
The mean field Hamiltonian \eqref{MFHamil} satisfies the following commutation   relation:
\begin{eqnarray}
[\hat{H}_{\rm MF}, \psi_m^{\dagger}(\boldsymbol{k}) \psi_n(\boldsymbol{k})]  &=& (\varepsilon_{m\bm{k}} - \varepsilon_{n \bm{k}} ) \psi_m^{\dagger}(\boldsymbol{k}) \psi_n(\boldsymbol{k}),
\end{eqnarray}
which leads to 
\begin{equation}
\hat{H}_{\rm MF} \psi_m^{\dagger}(\boldsymbol{k}) \psi_n(\boldsymbol{k}) |0\rangle \simeq 
\hat{H}_{\rm MF} \psi_m^{\dagger}(\boldsymbol{k}) \psi_n(\boldsymbol{k}) |0\rangle_{\rm MF} =
(\varepsilon_{m\bm{k}} - \varepsilon_{n \bm{k}} ) \psi_m^{\dagger}(\boldsymbol{k}) \psi_n(\boldsymbol{k})) |0\rangle_{\rm MF},
\end{equation}
where $|0\rangle$ is the true ground state.
Similarly,
\begin{eqnarray}
\hat{H}^{\nu}_{\rm MF} \psi_m^{\dagger}(\boldsymbol{k}) \psi_n(\boldsymbol{k}) |0\rangle \simeq 
(\varepsilon_{m\bm{k}} - \varepsilon_{n \bm{k}} )^{\nu} \psi_m^{\dagger}(\boldsymbol{k}) \psi_n(\boldsymbol{k})) |0\rangle_{\rm MF}.
\end{eqnarray}
for any integer number $\nu \geq 0$. In particular, this implies that
\begin{eqnarray}
 f(\hat{H}_{\rm MF}) \psi_m^{\dagger}(\boldsymbol{k}) \psi_n(\boldsymbol{k}) |0\rangle &\simeq &  \sum_{\nu=0}^{\infty} \frac{f^{(\nu)}(0)}{\nu!} 
 \hat{H}^{\nu}_{\rm MF} \psi_n(\boldsymbol{k}) |0\rangle_{\rm MF} = \sum_{\nu=0}^{\infty}
(\varepsilon_{m\bm{k}} - \varepsilon_{n \bm{k}})^{\nu}  \psi_m^{\dagger}(\boldsymbol{k}) \psi_n(\boldsymbol{k}) |0\rangle_{\rm MF}\\
&=& f(\varepsilon_{m\bm{k}} - \varepsilon_{n \bm{k}})  \psi_m^{\dagger}(\boldsymbol{k}) \psi_n(\boldsymbol{k}) |0\rangle_{\rm MF},
\end{eqnarray}
where $f$ is an analytic function. If we choose $f$ to be our mean field approximation of the resolvent, we obtain
\begin{equation}
\frac{1}{\hat{H}_{\rm MF}-\omega} \psi_m^{\dagger}(\boldsymbol{k}) \psi_n(\boldsymbol{k}) |0 \rangle \simeq
\frac{1}{\varepsilon_{m\bm{k}} - \varepsilon_{n \bm{k}} - \omega}
\psi_m^{\dagger}(\boldsymbol{k}) \psi_n(\boldsymbol{k})  |0 \rangle_{\rm MF}
\end{equation}
By using this relationship, we can rewrite the mean field Raman operators in the following way:
\begin{eqnarray}
 \hat{\bar{M}}_{R_1}^{\mu \nu}(\boldsymbol{q})  &= & \sum_{\boldsymbol{k} m n}  \sum_{\boldsymbol{k}^{\prime} m^{\prime} n^{\prime}} \psi_{m^{\prime}}^{\dagger}\left(\boldsymbol{k}^{\prime}\right) v_\nu^{m^{\prime} n^{\prime}}\left(\boldsymbol{k}^{\prime}, \boldsymbol{k}_f\right) \psi_{n^{\prime}}\left(\boldsymbol{k}^{\prime}\right) \frac{1}{ \varepsilon_{m\bm{k}} - \varepsilon_{n \bm{k}} - \omega_i} \psi_m^{\dagger}(\boldsymbol{k}) v_\mu^{m n}\left(\boldsymbol{k},-\boldsymbol{k}_i\right) \psi_n(\boldsymbol{k}),
 \\
\hat{\bar{M}}_{R_2}^{\mu \nu}(\boldsymbol{q}) &=&   \sum_{\boldsymbol{k}^{\prime} m^{\prime} n^{\prime}} \sum_{\boldsymbol{k} m n} \psi_m^{\dagger}(\boldsymbol{k}) v_\mu^{m n}\left(\boldsymbol{k},-\boldsymbol{k}_i\right) \psi_n(\boldsymbol{k}) 
\frac{1}{\varepsilon_{m'\bm{k}'} - \varepsilon_{n' \bm{k}'} + \omega_f} 
\psi_{m^{\prime}}^{\dagger}\left(\boldsymbol{k}^{\prime}\right) v_\nu^{m^{\prime} n^{\prime}}\left(\boldsymbol{k}^{\prime}, \boldsymbol{k}_f\right) \psi_{n^{\prime}}\left(\boldsymbol{k}^{\prime}\right).
\end{eqnarray}
Note that these are expected to be good approximations of the Raman operator only when they act on the ground state at zero temperature. 
The next goal is to extract the dominant contributions to the \emph{low-frequency} part of the Raman spectrum.
In the vicinity of the transition point induced by exciton condensation, as in the case of Sr$_3$Ir$_2$O$_7$, we only need contributions that create single particle-hole pair states.

Here, we determine the effective operator form for single particle-hole pair excitations.
Because the electronic current operator does not flip spins, the relevant excitations generated from $\hat{\bar{M}}_{R_1}^{\mu \nu}(\boldsymbol{q})$ acting on the ground state include: (i) $\bm{k}'=\bm{k}$, $m'=n'=m=(+,\sigma)$, and $n=(-,\sigma)$; (ii) $\bm{k}'=\bm{k}$, $m'=n'=(-,\bar{\sigma})$, $m=(+,\sigma)$, and $n=(-,\sigma)$, where $\bar{\sigma}$ is the spin opposite to $\sigma$; and (iii) $\bm{k}' \neq \bm{k}$, $m'=n'=(-,\sigma'), m=(+,\sigma)$, and $n=(-,\sigma)$.
The relevant cases for $\hat{\bar{M}}_{R_2}^{\mu \nu}(\boldsymbol{q})$ are similar.
Thus, the effective Raman operators become
\begin{align}
\hat{\bar{M}}_{R_1}^{\mu \nu}(\boldsymbol{q}) \rvert 0 \rangle_{\rm MF} &= \sum_{\boldsymbol{k} \sigma} \psi_{+\sigma}^{\dagger}\left(\boldsymbol{k}\right)  \frac{R^{\mu \nu}_\sigma \left( \bm{k}, \bm{k}_i, \bm{k}_f \right)}{ \varepsilon_{+\bm{k}} - \varepsilon_{-\bm{k}} - \omega_i} \psi_{-\sigma}(\boldsymbol{k}) \rvert 0 \rangle_{\rm MF} ,
 \\
 \hat{\bar{M}}_{R_2}^{\mu \nu}(\boldsymbol{q}) \rvert 0 \rangle_{\rm MF}  &= \sum_{\boldsymbol{k} \sigma} \psi_{+\sigma}^{\dagger}\left(\boldsymbol{k}\right) 
  \frac{R^{\nu \mu}_\sigma \left( \bm{k}, - \bm{k}_f, - \bm{k}_i \right)}{ \varepsilon_{+\bm{k}} - \varepsilon_{-\bm{k}} + \omega_f}  \psi_{-\sigma}(\boldsymbol{k}) \rvert 0 \rangle_{\rm MF},
\end{align}
where in an insulating phase,
\begin{align}
R^{\mu \nu}_\sigma \left( \bm{k}, \bm{k}_i, \bm{k}_f \right) &=\left(v_\nu^{+\sigma,+\sigma}\left(\boldsymbol{k}, \boldsymbol{k}_f\right) + v_\nu^{-\bar{\sigma},-\bar{\sigma}}\left(\boldsymbol{k}, \boldsymbol{k}_f \right)+ \sum_{\bm{k}'\neq \bm{k},\sigma'} v_\nu^{-\sigma',-\sigma'}\left(\bm{k}',\bm{k}_f \right)\right) v_\mu^{+\sigma,-\sigma}\left(\boldsymbol{k},-\boldsymbol{k}_i\right)
\\
&=\left(v_\nu^{+\sigma,+\sigma}\left(\boldsymbol{k}, \boldsymbol{k}_f\right) - v_\nu^{-\sigma,-{\sigma}}\left(\boldsymbol{k}, \boldsymbol{k}_f \right) \right) v_\mu^{+\sigma,-\sigma}\left(\boldsymbol{k},-\boldsymbol{k}_i\right).
\label{eq:R_ins}
\end{align}
We here used
\begin{align}
\sum_{\bm{k}',\sigma'} v_\nu^{-\sigma',-\sigma'}\left(\bm{k}',\bm{k}_f\right)=0. 
\end{align}
In a semimetallic phase, both bands are occupied in electron pockets (EP) and empty in hole pockets (HP), while the lower band is only occupied at the other wave vectors (Other).
Then, the coefficient $R^{\mu \nu}_\sigma \left( \bm{k}, \bm{k}_i, \bm{k}_f \right)$ is finite only at $\bm{k}\in {\rm Other}$ and given by
\begin{align}
R^{\mu \nu}_\sigma \left( \bm{k}, \bm{k}_i, \bm{k}_f \right) &=\left(v_\nu^{+\sigma,+\sigma}\left(\boldsymbol{k}, \boldsymbol{k}_f\right) + v_\nu^{-\bar{\sigma},-\bar{\sigma}}\left(\boldsymbol{k}, \boldsymbol{k}_f \right)+ \sum_{\bm{k}'\neq \bm{k},\bm{k}'\in {\rm Other}, \sigma'} v_\nu^{-\sigma',-\sigma'}\left(\bm{k}',\bm{k}_f \right) \right. \nonumber \\
& \qquad \left. + \sum_{\bm{k}'\in {\rm EP}, \sigma'} \left( v_\nu^{+\sigma',+\sigma'}\left(\bm{k}',\bm{k}_f \right) + v_\nu^{-\sigma',-\sigma'}\left(\bm{k}',\bm{k}_f \right) \right) \right) v_\mu^{+\sigma,-\sigma}\left(\boldsymbol{k},-\boldsymbol{k}_i\right)
\\
&=\left(v_\nu^{+\sigma,+\sigma}\left(\boldsymbol{k}, \boldsymbol{k}_f\right) - v_\nu^{-\sigma,-{\sigma}}\left(\boldsymbol{k}, \boldsymbol{k}_f \right) + \sum_{\bm{k}'\in {\rm EP}, \sigma'} v_\nu^{+\sigma',+\sigma'}\left(\bm{k}',\bm{k}_f \right) - \sum_{\bm{k}'\in {\rm HP}, \sigma'} v_\nu^{-\sigma',-\sigma'}\left(\bm{k}',\bm{k}_f \right) \right) v_\mu^{+\sigma,-\sigma}\left(\boldsymbol{k},-\boldsymbol{k}_i\right).
\label{eq:R_sm}
\end{align}
As mentioned above, the momenta of photons can be safely set to $\bm{k}_{i},\bm{k}_{f}=\bm{0}$.
Because the hopping amplitude is independent of the spin flavor in the $xy$ plane, we obtain
\begin{align}
v_\mu^{+\uparrow,-\uparrow}\left(\boldsymbol{k},\boldsymbol{0}\right) &= - v_\mu^{+\downarrow,-\downarrow}\left(\boldsymbol{k},\boldsymbol{0}\right),\\
v_\mu^{+\uparrow,+\uparrow}\left(\boldsymbol{k},\boldsymbol{0}\right) &= v_\mu^{+\downarrow,+\downarrow}\left(\boldsymbol{k},\boldsymbol{0}\right),\\
v_\mu^{-\uparrow,-\uparrow}\left(\boldsymbol{k},\boldsymbol{0}\right) &= v_\mu^{-\downarrow,-\downarrow}\left(\boldsymbol{k},\boldsymbol{0}\right).
\end{align}
We can rewrite the Raman operators as
\begin{eqnarray}
 \hat{\bar{M}}_{R_1}^{\mu \nu}(\boldsymbol{0}) \rvert 0 \rangle_{\rm MF}  &= & \sum_{\boldsymbol{k}}
 \frac{R^{\mu \nu}_\uparrow \left( \bm{k}, \bm{k}_i, \bm{k}_f \right)}{ \varepsilon_{+\bm{k}} - \varepsilon_{- \bm{k}} - \omega_i}  \left(\psi_{+\uparrow}^{\dagger}\left(\boldsymbol{k}\right) \psi_{-\uparrow}(\boldsymbol{k}) - \psi_{+\downarrow}^{\dagger}\left(\boldsymbol{k}\right) \psi_{-\downarrow}(\boldsymbol{k}) \right) \rvert 0 \rangle_{\rm MF},
 \\
  \hat{\bar{M}}_{R_2}^{\mu \nu}(\boldsymbol{0}) \rvert 0 \rangle_{\rm MF}  &= & \sum_{\boldsymbol{k}}  \frac{R^{\nu \mu}_\uparrow \left( \bm{k}, - \bm{k}_f, - \bm{k}_i \right)}{ \varepsilon_{+\bm{k}} - \varepsilon_{-\bm{k}} + \omega_f} \left(\psi_{+\uparrow}^{\dagger}\left(\boldsymbol{k}\right) \psi_{-\uparrow}(\boldsymbol{k}) - \psi_{+\downarrow}^{\dagger}\left(\boldsymbol{k}\right) \psi_{-\downarrow}(\boldsymbol{k}) \right) \rvert 0 \rangle_{\rm MF}.
\end{eqnarray}
Using the unitary transformation \eqref{eq:eigenvec}, the operator reads
\begin{equation}
 \hat{\bar{M}}^{\mu \nu}_R (\bm{0}) \rvert 0 \rangle_{\rm MF}=\sum_{\bm{k}}\bm{c}_{\bm{k}}^{\dagger}M^{\mu \nu}_{R,\bm{k}}\bm{c}_{\bm{k}} \rvert 0 \rangle_{\rm MF}, \label{eq:Mmf}
\end{equation}
where the vertex function is
\begin{equation}
M^{\mu \nu}_{R,\bm{k}}=\left(R_{1,\bm{k}}^{\mu\nu} + R_{2,\bm{k}}^{\mu\nu} \right)\left[\left(u_{\bm{k}}\tau_{+}+u_{\bm{k}}\tau_{-}\right)\otimes\sigma_{0}+\left(v_{\bm{k}}\tau_{+}-v_{\bm{k}}\tau_{-}+w_{\bm{k}}\tau_{z}\right)\otimes\sigma_{z}\right] \label{eq:M_F}
\end{equation}
with 
\begin{align}
R_{1,\bm{k}}^{\mu\nu} & =\frac{\left(v_\nu^{+\uparrow,+\uparrow}\left(\boldsymbol{k}, \boldsymbol{k}_f\right) - v_\nu^{-\uparrow,-{\uparrow}}\left(\boldsymbol{k}, \boldsymbol{k}_f \right) \right) v_\mu^{+\uparrow,-\uparrow}\left(\boldsymbol{k},-\boldsymbol{k}_i\right)}{\varepsilon_{+,\bm{k}}-\varepsilon_{-,\bm{k}}-\omega_{i}}, \label{eq:R1}\\
R_{2,\bm{k}}^{\mu\nu} & =\frac{\left(v_\mu^{+\uparrow,+\uparrow}\left(\boldsymbol{k}, -\boldsymbol{k}_i\right) - v_\mu^{-\uparrow,-{\uparrow}}\left(\boldsymbol{k}, -\boldsymbol{k}_i \right) \right) v_\nu^{+\uparrow,-\uparrow}\left(\boldsymbol{k},\boldsymbol{k}_f\right)}{\varepsilon_{+,\bm{k}}-\varepsilon_{-,\bm{k}}+\omega_{f}}, \label{eq:R2}\\
u_{\bm{k}} & = - \frac{1}{4}\left( x_{\uparrow \bm{k}} (1+z_{\bm{k}}) - \bar{x}_{\uparrow \bm{k}} (1-z_{\bm{k}}) \right),\\
v_{\bm{k}} & = - \frac{1}{4}\left( x_{\uparrow \bm{k}} (1+z_{\bm{k}}) + \bar{x}_{\uparrow \bm{k}} (1-z_{\bm{k}}) \right),\\
w_{\bm{k}} & =\frac{1}{2}\sqrt{1-z_{\bm{k}}^{2}},
\end{align}
in an insulating phase. In a semimetallic phase, the numerators of $R_{1,\bm{k}}^{\mu\nu}$ and $R_{2,\bm{k}}^{\mu\nu}$ are modified according to the difference between Eqs.~\eqref{eq:R_ins} and \eqref{eq:R_sm}.
Here $\tau_{\pm}=(\tau_{x}\pm i\tau_{y})/2$ and $\tau_{x,y,z}$ are Pauli
matrices in the sublattice space. 
Note that the intensity of the resonant contribution is significantly enhanced when the band energy gap is close to the incident photon energy: $\varepsilon_{+,\bm{k}}-\varepsilon_{-,\bm{k}} \simeq \omega_{i}$ in Eqs.~\eqref{eq:R1} and \eqref{eq:R2}.

In total, summing up the nonresonant and resonant contributions, the effective Raman operator acting on the ground state is given by 
\begin{equation}
 \hat{\bar{M}}^{\mu \nu} (\bm{0}) \rvert 0 \rangle_{\rm MF}=\sum_{\bm{k}}\bm{c}_{\bm{k}}^{\dagger} M^{\mu\nu}_{\bm{k}} \bm{c}_{\bm{k}} \rvert 0 \rangle_{\rm MF}, \label{eq:Mtotal}
\end{equation}
where
\begin{equation}
    M^{\mu\nu}_{\bm{k}} = M^{\mu \nu}_{N,\bm{k}} + M^{\mu \nu}_{R,\bm{k}} \label{eq:Mmat}
\end{equation}
defined by Eqs.~\eqref{eq:M_N0} and \eqref{eq:M_F}.

\section{Random phase approximation}
\begin{figure}[tb]
\centering
\includegraphics[scale=0.15]{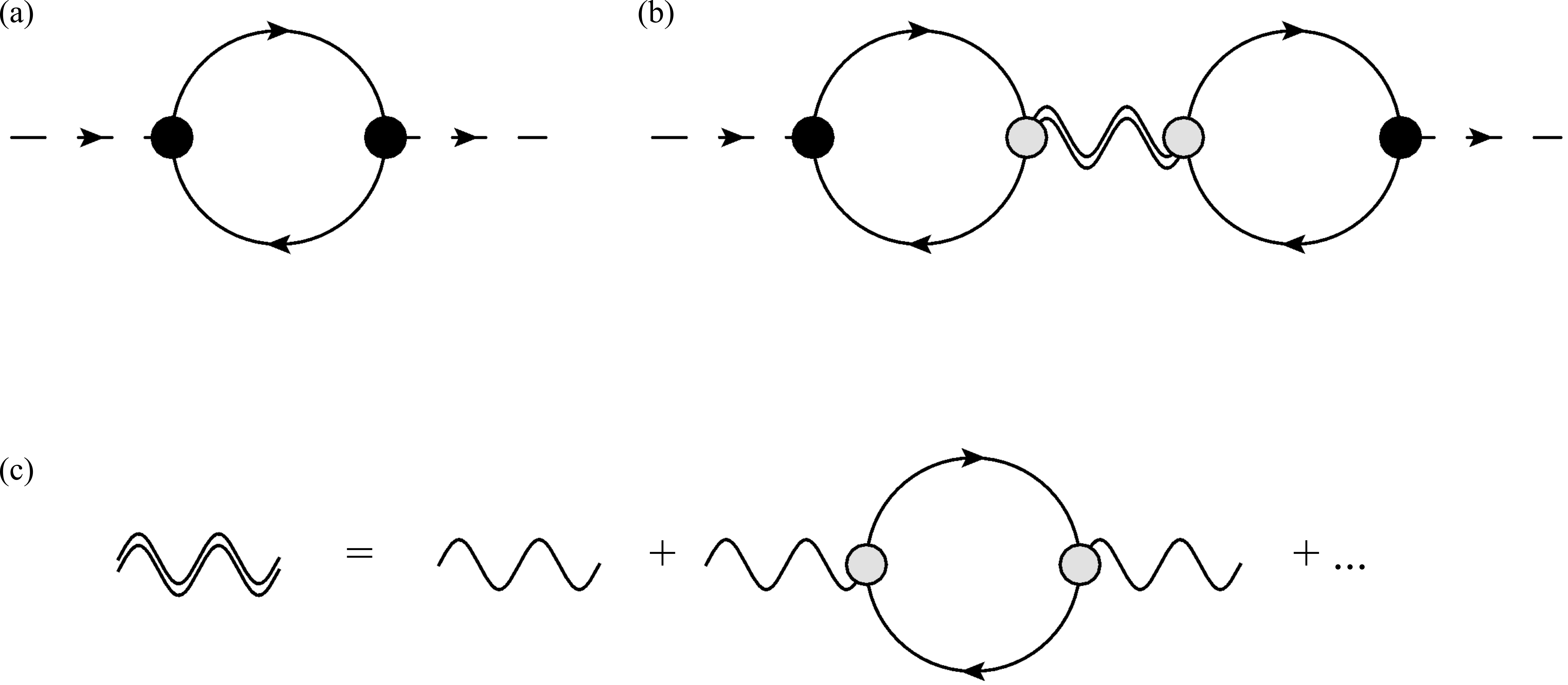}\caption{Susceptibility (a) $\sigma_{0}(\omega)$ in Eq. (\ref{eq:sig0}) and
(b) $\sigma_{1}(\omega)$ in Eq. (\ref{eq:sig1}). (c) The dressed
RPA interaction vertex. 
The black solid circles represent the Raman operator vertex
in Eq. (\ref{eq:Mmat}), and the gray open circles represent the interaction vertex in Eq. (\ref{eq:vint}). }
\label{fig:rpa}
\end{figure}

The Raman scattering intensity is proportional to the Fourier transform of the two-point correlation function
\begin{equation}
R^{\bm{e}^i \bm{e}^f}(\omega) = \int_{-\infty}^{\infty} dte^{i\omega t}\langle {\bm e}^i_{\mu} {\bm e}^f_{\nu} \hat{\bar{M}}^{\mu \nu \dagger} (t)  {\bm e}^i_{\alpha} {\bm e}^f_{\beta} \hat{\bar{M}}^{\alpha \beta}(0)\rangle,
\end{equation}
where we are adopting the convention of summation of repeated Greek indices.
It is related to the imaginary
part of the dynamic susceptibility of the Raman operator (fluctuation-dissipation theorem),
\begin{equation}
R^{\bm{e}^i \bm{e}^f}(\omega)=-2\text{Im}\sigma^{\bm{e}^i \bm{e}^f}(\omega),
\end{equation}
where
\begin{equation}
\sigma^{\bm{e}^i \bm{e}^f}(\omega)=-i\int_{0}^{\infty}dte^{i\omega t}\langle [  {\bm e}^i_{\mu} {\bm e}^f_{\nu} \hat{\bar{M}}^{\mu \nu \dagger}(t),  {\bm e}^i_{\alpha} {\bm e}^f_{\beta} \hat{\bar{M}}^{\alpha \beta}(0)]\rangle.
\end{equation}
To obtain $\sigma^{\bm{e}^i \bm{e}^f}(\omega)$, we first calculate the corresponding
correlation function on the imaginary axis:
\begin{equation}
\sigma^{\bm{e}^i \bm{e}^f}(i\omega_{n})=-\int_{0}^{\beta}d\tau e^{i\omega_{n}\tau}\langle  {\bm e}^i_{\mu} {\bm e}^f_{\nu} \hat{\bar{M}}^{\mu \nu \dagger}(\tau)  {\bm e}^i_{\alpha} {\bm e}^f_{\beta} \hat{\bar{M}}^{\alpha \beta}(0)\rangle.
\end{equation}
For a noninteracting gas of fermions, this correlator is given by
\begin{eqnarray}
-\sigma^{\bm{e}^i \bm{e}^f}_{0}(i\omega_{n})&=&\frac{1}{{\cal N}_{u}}\sum_{\bm{k}}\sum_{k,l}\frac{n_{F}(\varepsilon_{l\sigma^{\prime}}(\bm{k}))-n_{F}(\varepsilon_{k\sigma}(\bm{k}))}{i\omega_{n}-(\varepsilon_{l\sigma^{\prime}}(\bm{k})-\varepsilon_{k\sigma}(\bm{k}))} \nonumber \\
& \times &{\rm Tr}\left[P_{l\sigma^{\prime}}(\bm{k})  {\bm e}^i_{\alpha} {\bm e}^f_{\beta} M^{\alpha \beta}_{\bm{k}}P_{k\sigma}(\bm{k})  {\bm e}^i_{\mu} {\bm e}^f_{\nu} M^{\mu \nu \dagger}_{\bm{k}} \right],
\label{eq:sig0}
\end{eqnarray}
where
${\cal N}_u$ is the number of unit cells,
$n_F(\varepsilon)=\frac{1}{1+e^{\beta (\varepsilon - \mu)}}$ is the Fermi distribution with the chemical potential $\mu$, $P_{n\sigma}(\bm{k})=X_{n\sigma}(\bm{k})X_{n\sigma}^{\dagger}(\bm{k})$
is the projector on the eigenstate $X_{n\sigma}(\bm{k})$ (column
vector) of the mean-field Hamiltonian given by Eq. \eqref{eq:eigenvec}, and $M^{\mu\nu}_{\bm{k}}$ is the matrix form of the effective Raman operator defined by Eq.~\eqref{eq:Mmat}. 
Figure \ref{fig:rpa} (a) shows $\sigma_{0}(i\omega_{n})$ diagrammatically.

In our system, the fermions interact with each other through the following
on-site interaction 
\begin{equation}
{\cal H}_{I}=\frac{1}{2}\sum_{\bm{r}}\sum_{\{\sigma_{i}\}}V_{\sigma_{1}\sigma_{4};\sigma_{2}\sigma_{3}}c_{\sigma_{1},\bm{r}}^{\dagger}c_{\sigma_{2},\bm{r}}^{\dagger}c_{\sigma_{3},\bm{r}}c_{\sigma_{4},\bm{r}},
\end{equation}
where the interaction vertex takes three equivalent forms
\begin{align}
V_{\sigma_{1}\sigma_{4};\sigma_{2}\sigma_{3}} & =\frac{1}{2}U\left(\sigma_{\sigma_{1}\sigma_{4}}^{0}\sigma_{\sigma_{2}\sigma_{3}}^{0}-\sigma_{\sigma_{1}\sigma_{4}}^{z}\sigma_{\sigma_{2}\sigma_{3}}^{z}\right) \label{eq:vint}\\
 & =\frac{1}{2}U\left(\sigma_{\sigma_{1}\sigma_{4}}^{0}\sigma_{\sigma_{2}\sigma_{3}}^{0}-2\sigma_{\sigma_{1}\sigma_{4}}^{+}\sigma_{\sigma_{2}\sigma_{3}}^{-}\right)\\
 & =\frac{1}{2}U\left(\sigma_{\sigma_{1}\sigma_{4}}^{0}\sigma_{\sigma_{2}\sigma_{3}}^{0}-2\sigma_{\sigma_{1}\sigma_{4}}^{-}\sigma_{\sigma_{2}\sigma_{3}}^{+}\right).
\end{align}
Including the contributions from the interaction, the total $\sigma(\omega)$
has two parts:
\begin{equation}
\sigma(\omega)=\sigma_{0}(\omega)+\sigma_{1}(\omega).
\end{equation}
In the random phase approximation (RPA), $\sigma_{1}(\omega)$ is
obtained by summing up the bubble diagrams (see Fig. \ref{fig:rpa}
(b)). 
There are two types of bubbles. 
The first type connects an external Ramnan vertex [Eq. \eqref{eq:Mmat}] and an interaction
vertex [Eq. (\ref{eq:vint})], we call the \textit{external loop}.
It describes the correlator of the Raman and spin/charge operators
for a noninteracting system.
The right external loop in Fig.~\ref{fig:rpa} (b) is given by
\begin{eqnarray}
\Pi_{\gamma}^{a}(i\omega_{n})&=&\frac{1}{{\cal N}_{u}}\sum_{\bm{k}}\sum_{k,l}\frac{n_{F}(\varepsilon_{l\sigma^{\prime}}(\bm{k}))-n_{F}(\varepsilon_{k\sigma}(\bm{k}))}{i\omega_{n}-(\varepsilon_{l\sigma^{\prime}}(\bm{k})-\varepsilon_{k\sigma}(\bm{k}))} \nonumber \\
&\times&{\rm Tr}\left[P_{l\sigma^{\prime}}(\bm{k})\Sigma^{\gamma,a}P_{k\sigma}(\bm{k})  {\bm e}^i_{\mu} {\bm e}^f_{\nu}M^{\mu \nu \dagger}_{\bm{k}}\right]
\end{eqnarray}
where
$\Sigma^{\gamma,a}\equiv\frac{1}{2}(\tau_{0}+\epsilon_\gamma \tau_{z})\otimes\sigma^{a}$ with $\epsilon_\gamma=1(-1)$ for sublattice $\gamma=\mathcal{A}(\mathcal{B})$ and $a=0,z,+,-$.
The second type of bubble connects two interaction
vertices [Eq. (\ref{eq:vint})], we call the \textit{internal loop}. 
It gives the charge/spin susceptibility
of a noninteracting system:
\begin{eqnarray}
\chi_{\gamma_1\gamma_2}^{ab}(i\omega_{n})&=&\frac{1}{{\cal N}_{u}}\sum_{\bm{k}}\sum_{k,l}\frac{n_{F}(\varepsilon_{l\sigma^{\prime}}(\bm{k}))-n_{F}(\varepsilon_{k\sigma}(\bm{k}))}{i\omega_{n}-(\varepsilon_{l\sigma^{\prime}}(\bm{k})-\varepsilon_{k\sigma}(\bm{k}))} \nonumber \\
&\times &{\rm Tr}\left[P_{l\sigma^{\prime}}(\bm{k})\Sigma^{\gamma_1,a}P_{k\sigma}(\bm{k})\Sigma^{\gamma_2,b}\right], \label{eq:chi}
\end{eqnarray}
where $\gamma_1,\gamma_2$ are the sublattice index.
These internal loops can be summed, resulting in a dressed interaction
vertex (see Fig. \ref{fig:rpa} (c)). Given the U(1) invariance of
${\cal H}$ and the ground state, the fluctuations of the charge,
transverse spin, and longitudinal spin sectors are decoupled from
each other. In other words, we get from this symmetry argument that
$\chi^{ab}\equiv\chi^{00}\oplus\chi^{zz}\oplus\chi^{-+}\oplus\chi^{+-}$.
Thus, we get
\begin{align}
-\sigma_{1}(i\omega_{n}) & =\left(\Pi^{0*}(-i\omega_{n})\right)^{T}\tilde{U}_{00}(\bm{0},i\omega_{n})\Pi^{0}(i\omega_{n})+\left( \Pi^{z*}(-i\omega_{n})\right)^{T}\tilde{U}_{zz}(\bm{0},i\omega_{n})\Pi^{z}(i\omega_{n})\nonumber \\
 & +\left(\Pi^{-*}(-i\omega_{n})\right)^{T}\tilde{U}_{+-}(\bm{0},i\omega_{n})\Pi^{-}(i\omega_{n})+\left(\Pi^{+*}(-i\omega_{n})\right)^{T}\tilde{U}_{-+}(\bm{0},i\omega_{n})\Pi^{+}(i\omega_{n}).\label{eq:sig1}
\end{align}
where the dressed interaction vertices are given by
\begin{align}
\tilde{U}_{00} & =\frac{-\frac{1}{2}U}{1+\frac{1}{2}U\chi^{00}(\bm{q},i\omega_{n})},\\
\tilde{U}_{zz} & =\frac{\frac{1}{2}U}{1-\frac{1}{2}U\chi^{zz}(\bm{q},i\omega_{n})},\\
\tilde{U}_{+-} & =\frac{U}{1-U\chi^{+-}(\bm{q},i\omega_{n})},\\
\tilde{U}_{-+} & =\frac{U}{1-U\chi^{-+}(\bm{q},i\omega_{n})}.
\end{align}
Due to the symmetry of the Raman operator considered in this work, only the longitudinal spin channel ($zz$) gives a finite contribution to $\sigma_1(\omega)$.

\end{document}